
%
%
%
%
%
%
\input harvmac
%
%
%
%
%
%
\ifx\answ\bigans
\else
\output={
  \almostshipout{\leftline{\vbox{\pagebody\makefootline}}}\advancepageno
}
\fi
%
%
%

%
%

%
%
\def\UCSD#1#2{\noindent#1\hfill #2%
\bigskip\supereject\global\hsize=\hsbody%
\footline={\hss\tenrm\folio\hss}}
%
%
\def\abstract#1{\centerline{\bf Abstract}\nobreak\medskip\nobreak\par #1}
%
%
%
%
\edef\tfontsize{ scaled\magstep3}
 \tfontsize  \tfontsize
 \tfontsize \font\titlei=cmmi10 \tfontsize
\font\titleis=cmmi7 \tfontsize \font\titleiss=cmmi5 \tfontsize
\font\titlesy=cmsy10 \tfontsize \font\titlesys=cmsy7 \tfontsize
\font\titlesyss=cmsy5 \tfontsize  \tfontsize
\skewchar\titlei='177 \skewchar\titleis='177 \skewchar\titleiss='177
\skewchar\titlesy='60 \skewchar\titlesys='60 \skewchar\titlesyss='60
%
%
%
%
%
\def\inv{^{\raise.15ex\hbox{${\scriptscriptstyle -}$}\kern-.05em 1}}
\def\lbar{{\lower.35ex\hbox{$\mathchar'26$}\mkern-10mu\lambda}} 

%
%
%
%
\def\slash#1{\rlap{$#1$}/} 
\def\dsl{\,\raise.15ex\hbox{/}\mkern-13.5mu D} 
\def\delsl{\raise.15ex\hbox{/}\kern-.57em\partial}
\def\Ksl{\hbox{/\kern-.6000em\rm K}}
\def\Asl{\hbox{/\kern-.6500em \rm A}}
\def\Dsl{\hbox{/\kern-.6000em\rm D}} 
\def\Qsl{\hbox{/\kern-.6000em\rm Q}}
\def\gradsl{\hbox{/\kern-.6500em$\nabla$}}
%
%
\def\lspace{\ifx\answ\bigans{}\else\qquad\fi}
\def\lbspace{\ifx\answ\bigans{}\else\hskip-.2in\fi} 
%
%
\def\boxeqn#1{\vcenter{\vbox{\hrule\hbox{\vrule\kern3pt\vbox{\kern3pt
        \hbox{${\displaystyle #1}$}\kern3pt}\kern3pt\vrule}\hrule}}}
%
%
\def\mbox#1#2{\vcenter{\hrule \hbox{\vrule height#2in
\kern#1in \vrule} \hrule}}
%
%
%
%

   \def\CL{{\cal L}}

%
%
%
%
%

%

\def\bar#1{\overline{#1}}
\def\vev#1{\left\langle #1 \right\rangle}

\def\darr#1{\raise1.5ex\hbox{$\leftrightarrow$}\mkern-16.5mu #1}

%
%
\def\half{{\textstyle{1\over2}}} 
\def\frac#1#2{{\textstyle{#1\over #2}}} 
%
%
%
%

\def\Tr{\mathop{\rm Tr}}

%
%
%
%

%
%
\def\ltap{\ \raise.3ex\hbox{$<$\kern-.75em\lower1ex\hbox{$\sim$}}\ }
\def\gtap{\ \raise.3ex\hbox{$>$\kern-.75em\lower1ex\hbox{$\sim$}}\ }
\def\gl{\ \raise.5ex\hbox{$>$}\kern-.8em\lower.5ex\hbox{$<$}\ }
\def\roughly#1{\raise.3ex\hbox{$#1$\kern-.75em\lower1ex\hbox{$\sim$}}}
%
%
\def\ie{\hbox{\it i.e.}}        \def\etc{\hbox{\it etc.}}
\def\eg{\hbox{\it e.g.}}        
\def\etal{\hbox{\it et al.}}

\def\np#1#2#3{Nucl. Phys. B{#1} (#2) #3}
\def\pl#1#2#3{Phys. Lett. {#1}B (#2) #3}
\def\prl#1#2#3{Phys. Rev. Lett. {#1} (#2) #3}
\def\physrev#1#2#3{Phys. Rev. {#1} (#2) #3}
\def\ap#1#2#3{Ann. Phys. {#1} (#2) #3}

\relax

\def\ref{${}^{\the\refno)}$\nref}
\def\nref#1{\xdef#1{\the\refno)}\writedef{#1\leftbracket#1}%
\ifnum\refno=1\immediate\openout\rfile=refs.tmp\fi
\global\advance\refno by1\chardef\wfile=\rfile\immediate
\write\rfile{\noexpand\item{#1\ }\reflabeL{#1\hskip.31in}\pctsign}\findarg}
\hsize 17truecm
\hoffset -0.3truecm
\vsize 24.94truecm
\voffset -1.25truecm
\ifx\epsfbox\notincluded\message{(NO epsf.tex, FIGURES WILL BE IGNORED)}
\def\insertfig#1#2#3{}
\else\message{(FIGURES WILL BE INCLUDED)}
\def\insertfig#1#2#3{
\midinsert\centerline{\epsffile{#3}}
{\baselineskip=0.45truecm{
\centerline{{#1}}\centerline{{#2}}
}}
\endinsert}\fi
\phantom{a}
\vskip 4.85truecm
\centerline{{CHIRAL PERTURBATION THEORY}}
\vskip 0.75truecm
{\baselineskip=0.45truecm
\centerline{Aneesh V.~Manohar}
\vskip 0.5truecm
\centerline{Department of Physics, University of California
San Diego, La Jolla, CA 92093, USA}
\smallskip
\centerline{and}
\smallskip
\centerline{CERN TH Division, CH-1211 Geneva 23, Switzerland}
\vfill}
\vskip -2.1truecm
\noindent
ABSTRACT
\vskip 0.55truecm
{\baselineskip=0.45truecm
\noindent
The basic ideas and power counting rules of chiral perturbation theory are
discussed. The formalism of velocity dependent fields for baryons, and for
hadrons containing a heavy quark, is explained.
\vfill}
\UCSD{\vbox{\hbox{UCSD/PTH
93-12}\hbox{hep-ph/9305298}}}{May 1993}
\eject

\def\sulr{SU(3)_L\times SU(3)_R}
\def\lchi{\Lambda_\chi}

Classically, the QCD Lagrangian with three massless quarks
has a global $U(3)_L\times U(3)_R$ symmetry under
which the left- and right-handed quark fields transform as the fundamental
representations of $U(3)_L$ and $U(3)_R$ respectively. As is well
known, there is an anomaly in the quantum theory so that only the
$SU(3)_L\times SU(3)_R \times U(1)_V$ subgroup of
$U(3)_L\times U(3)_R$ is a good symmetry; the $U(1)_A$ axial
symmetry is not preserved by quantum corrections.
(In this talk, I will not worry about global properties, so $U(3)$
will be considered to be equivalent to $SU(3)\times U(1)$.)
The spectrum of physical states does not form a representation of
$SU(3)_L\times SU(3)_R \times U(1)_V$, but only of the
$SU(3)_V\times U(1)_V$ vector subgroup. For example, the baryon
spectrum is not parity doubled, but is an $SU(3)_V$ octet with baryon number
one. This implies that the
$SU(3)_L\times SU(3)_R \times U(1)_V$ of the Lagrangian is
spontaneously broken down to $SU(3)_V\times U(1)_V$ at a scale
$\lchi\sim1$ GeV, so there are eight Goldstone bosons, the $\pi$'s, $K$'s
and $\eta$. The interactions of the Goldstone bosons at energies much
smaller than $\lchi$ can be calculated using an effective
Lagrangian. The light quarks $u$, $d$, and $s$
can be treated as approximately
massless and have masses that are small compared with $\lchi$,
whereas the heavy quarks $c$, $b$ and $t$ are those whose masses are
large compared with $\lchi$. The heavy quarks are integrated out
of the low energy effective theory, and will be neglected in this
discussion.

The chiral flavor symmetry is spontaneously broken by the vacuum
condensate
\eqn\cond{
\vev{\bar\psi_{jR}\ \psi_{iL}}\equiv a \ \Sigma_{ij}\ ,
}
where $a$ is the magnitude of the condensate, and $\Sigma$ is the
direction of the condensate in flavor space. $\Sigma$ is normalized so that
$\Sigma\Sigma^\dagger =1$.  All orientations of $\Sigma$ are equivalent vacuum
states, and can be rotated into each other by applying an $\sulr$
transformation, under which
\eqn\trans{
\Sigma \rightarrow L\Sigma R^\dagger\ .
}
The low energy excitations of the theory are ones in which the system is
in a local vacuum configuration at each point in space, but where the
orientation of the vacuum varies slowly. These configurations can be
described by eq.~\cond\ where the field $\Sigma$ which describes the
orientation of the vacuum is now a slowly varying function of position. The
excitation energy can be made arbitrarily small by making the changes in
$\Sigma$ over arbitrarily large length scales. These low energy excitations are
 the massless Goldstone bosons.
Configurations in which the magnitude of the condensate $a$ changes with
position have finite energy, and can be neglected.

The low energy effective theory contains only the massless Goldstone boson
modes. This theory does not have infrared divergences because the Goldstone
boson couplings vanish at zero momentum. This is trivial to see; a zero
momentum Goldstone boson corresponds to a $\Sigma$ field that is constant in
space, \ie\ to one of the vacuum configurations of the theory. Thus a zero
momentum Goldstone boson is equivalent to nothing, and cannot interact.

QCD is a  complicated theory in which the origin of the symmetry breaking is
non-perturbative. There is no known way of explicitly deriving the low energy
QCD effective theory directly from the original QCD Lagrangian. One must
therefore write down the most general possible Lagrangian for the $\Sigma$
field of the effective theory, consistent with the chiral symmetries of the
original QCD Lagrangian, eq.~\trans, and with $C$, $P$ and $T$ invariance. The
parameters of this effective Lagrangian can in principle be computed from the
QCD theory, but in practice, they have to obtained by comparison with
experiment. There is no term in the effective Lagrangian with zero derivatives.
The only allowed term is proportional to $\Tr \Sigma\Sigma^\dagger$, and is a
constant, since $\Sigma\Sigma^\dagger=1$. This is a consequence of
zero-momentum Goldstone bosons being equivalent to the vacuum. The first term
in the effective Lagrangian has two derivatives, and is conventionally written
as
\eqn\ltwo{
\CL_2 = {f^2\over4} \Tr\partial_\mu\,\Sigma\partial^\mu\Sigma^\dagger\ ,
}
and is the only allowed term with two derivatives. The field $\Sigma$ is the
exponential of the Goldstone boson fields $\pi$ (the Goldstone boson fields
should be thought of as angular variables describing the vacuum rotation),
\eqn\sigdef{
\Sigma(x) = e^{2 i \pi(x)/f},\qquad \pi(x)=\pi^A(x) T^A,\ A=1,\ldots,8\ ,
}
and the $SU(3)$ generators are normalized to $\Tr T^AT^B$ = $\half\delta^{AB}$.
 Expanding eq.~\sigdef\ and substituting into eq.~\ltwo\ gives
\eqn\expnsn{
\CL_2=\Tr\partial_\mu\pi \partial^\mu\pi + {1\over 2 f^2}\Tr
\left[\pi,\partial_\mu \pi\right]^2 + \ldots \ .
}
The higher order terms in the expansion of eq.~\expnsn\ give non-linear
Goldstone boson interactions. All the interactions in eq.~\expnsn\ are
completely determined in terms of a single unknown parameter $f$, the pion
decay constant. Note that the $2\pi$, $4\pi$, $6\pi$, \etc\ terms are related
to each other by the spontaneously broken chiral symmetry. This is a general
result; a spontaneously broken symmetry relates processes involving different
numbers of Goldstone bosons.

The effective Lagrangian also contains terms involving more than two
derivatives. One possible term involving four derivatives is
\eqn\four{
\CL_4 = c\, {f^2\over 4} {1\over \lchi^2} \Tr
\partial_\mu\Sigma\,
\partial_\nu\Sigma^\dagger\,\partial^\mu\Sigma\,
\partial^\nu\Sigma^\dagger\ ,
}
where $c$ is an unknown dimensionless coefficient. The term has been normalized
relative to the lowest order term eq.~\ltwo\ by introducing a factor $\lchi$
for each additional derivative.  With this normalization convention, $c$ is
expected to be of order unity. $\lchi$ is the scale at which the derivative
expansion of the effective theory breaks down, and will be called the scale of
chiral symmetry breaking. In QCD, $\lchi$ is of order 1~GeV.

The effective Lagrangian gives an expansion for a multi-Goldstone boson
scattering amplitude in powers of momentum. For example, the $\pi-\pi$
scattering amplitude is obtained by expanding $\CL$ up to order $\pi^4$. The
leading order amplitude is of order $p^2/f^2$ from $\CL_2$, the first
correction is of order $c p^4/f^2\lchi^2$ from $\CL_4$, \etc\ One must also
consider loop graphs in the effective Lagrangian. These graphs are in general
infinite, and need to be renormalized. It is extremely useful to use a mass
independent renormalization scheme (such as dimensional regularization and
minimal subtraction) in an effective field theory. In such a scheme, the
dimensional scale $\mu$ introduced to regulate the graphs can only occur in the
form $\log \mu$. Consider, for example the $\pi$-$\pi$ scattering graph of
Figure 1, \insertfig{Figure 1}{One loop correction to $\pi-\pi$
scattering.}{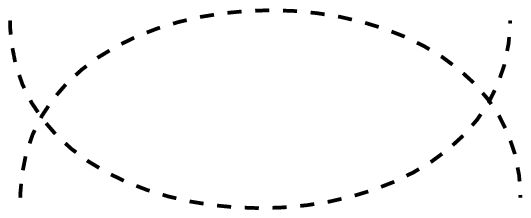}
\noindent where the two vertices in the graph are from $\CL_2$.
The form of the loop graph is
\eqn\loopform{
A = \int {d^4k\over (2\pi)^4} {p^2\over f^2}{p^2\over f^2}\left({1\over
p^2}\right)^2\ ,
}
where $p$ and $k$ are generic momenta. Each four pion vertex from $\CL_2$ is of
order $p^2/f^2$, the two pion propagators are of order $1/p^2$, and $k$ denotes
the loop momentum. By dimensional analysis, the loop integral must have the
form
\eqn\aresult{
A \sim {1\over 16 \pi^2 f^4} p^4 \log(p/\mu)\ ,
}
where the $16\pi^2$ is the standard factor that comes from loop diagrams. Note
that the form of the amplitude is determined  (up to logarithms) by dimensional
analysis, since there can be no powers of $\mu$ in the Feynman graph. Thus a
one loop graph with vertices from $\CL_2$ gives an amplitude which is of the
same order in momentum as that from $\CL_4$. The $\mu$ dependence of $A$ in
eq.~\aresult\ is canceled by the $\mu$ dependence of the various coefficients
(such as $c$) in $\CL_4$. A change in $\mu$ of order one produces a shift in
the coefficient $c$ of order $\lchi^2/(16\pi^2 f^2)$. This suggests that one
should define the scale $\lchi$ to be $4\pi f\sim 1$ GeV \ref\mg{A. Manohar and
H. Georgi, \np{234}{1984}{189}}, so that $c$ is then expected to be of order
one. With this prescription, we see that the order $p^4$ amplitude is given by
a tree graph from $\CL_4$ and a one loop graph from $\CL_2$. This scheme can
shown to be consistent to all orders in the derivative and loop
expansions${}^{\mg}$.  In general, one can show\ref\weinberg{S. Weinberg,
Physica 96A (1979) 327} that an $L$ loop graph with $m_k$ vertices from
$\CL_{k}$ contributes an amplitude which is of the same order of magnitude as a
contribution from $\CL_m$, where
\eqn\power{
m - 2 = \sum_{k\ge 2} m_k(k -2 ) + 2 L\ .
}
Note that $k\ge2$, since the Lagrangian begins at order $p^2$.
Terms of order $p^4$ can come from $L=1$, with $m_k=0$ for $k>2$, or from $L=0$
and $m_k=1$ for $k=4$ and $m_k=0$ for $k>4$, \ie\ one-loop graphs from the
lowest order Lagrangian $\CL_2$ or tree graphs from the lowest order Lagrangian
with one insertion of $\CL_4$. This is precisely what was obtained by an
explicit computation for the $\pi-\pi$ scattering amplitude. Similarly, terms
of order $p^6$ can come from two-loop graphs using $\CL_2$, one-loop graphs
with one insertion of $\CL_4$, or tree graphs with two insertions of $\CL_4$ or
with one insertion of $\CL_6$.

There is a very important consequence of eq.~\power. The low energy effective
theory gives a scattering amplitude as a power series expansion in $p/\lchi$.
If we work to a fixed order in the momentum expansion, then only a finite
number of terms in the effective Lagrangian are relevant. Thus, the theory has
only a finite number of parameters, and is equivalent to a renormalizable
theory. The number of parameters can be very large. However, since the
effective Lagrangian relates processes involving different numbers of Goldstone
bosons, there are (in principle) an infinite number of experimental amplitudes
that can be fit to a finite number of parameters. This method has been used
with great success in QCD by Gasser and Leutwyler\ref\gl{J. Gasser and H.
Leutwyler, \np{250}{1985}{465}\semi J. Gasser and H. Leutwyler,
\np{250}{1985}{517}\semi J. Gasser and H. Leutwyler, \ap{158}{1984}{142}}.

The QCD theory also has other states in the theory which are not Goldstone
bosons, the baryons, $\rho$ mesons, \etc, which are generically called matter
fields. The general theory of the interaction of matter fields with Goldstone
bosons was worked out by Callan, Coleman, Wess and Zumino\ref\ccwz{S.~Coleman,
J.~Wess and B.~Zumino, Phys. Rev. 177 (1969) 2239\semi C.~Callan, S.~Coleman,
J.~Wess and B.~Zumino, Phys. Rev. 177 (1969) 2247}. I will concentrate here on
the baryon octet ($p$, $n$, $\Lambda$ and $\Sigma$), but the method is general
and applies to the other cases as well. Matter fields transform as definite
representations of the unbroken $SU(3)_V$ symmetry group, but do not form
definite representations of broken chiral $\sulr$. The CCWZ result is that one
should pick any transformation property for the matter fields under chiral
$\sulr$ that reduces to the correct transformation property under the unbroken
subgroup $SU(3)_V$, and use the most general possible Lagrangian consistent
with the symmetries. Different choices of the $\sulr$ representation can be
shown to lead to the same $S$-matrix. For example, in the case of the octet
baryons, one can choose
$$
\eqalign{
B \rightarrow L B L^\dagger,&\qquad{\rm or}\qquad B \rightarrow R B
R^\dagger,\cr
{\rm or}\qquad B\rightarrow L B R^\dagger, &\qquad{\rm or}\qquad B \rightarrow
R B L^\dagger\ .\cr
}
$$
All of these reduce to the required octet transformation law $B\rightarrow V B
V^\dagger$ for vector transformations under which $L=R=V$. One commonly used
choice is to use different transformation laws for the left and right handed
baryon fields,
$$
B_L \rightarrow L B_L L^\dagger,\qquad\qquad B_R\rightarrow R B_R R^\dagger.
$$
A more useful choice is to define the field $\xi$
\eqn\xidef{
\xi = e^{i\pi/f},\qquad\qquad \Sigma=\xi^2\ ,
}
which has the $\sulr$ transformation law
\eqn\xitrans{
\xi(x) \rightarrow L\, \xi(x) \, U(x)^\dagger = U(x)\, \xi(x)\, R^\dagger\ .
}
This equation implicitly defines the matrix $U$ in terms of $L$, $R$, and
$\xi$.
Note that $U$ depends on $x$ through its dependence on $\xi$, even for a global
$\sulr$ transformation. The transformation law for the baryon fields is chosen
to be
\eqn\bary{
B \rightarrow U B U^\dagger\ .
}
Under vector transformations for which $L=R=V$, eq.~\xitrans\ implies that
$U=V$, so that eq.~\bary\ reduces to the required octet transformation law
$B\rightarrow V B V^\dagger$. The advantage of the $U$ basis is that the
effective Lagrangian for the baryon fields has only terms in which the pion is
explicitly derivatively coupled, so that it is simple to take the low-momentum
limit.

The effective theory for baryons is conveniently written in terms of the fields
\eqn\vadef{\eqalign{
V_\mu =& \half\left(\xi\partial^\mu\xi^\dagger +
\xi^\dagger\partial^\mu\xi\right) = {1\over f^2}\left[\pi,\partial_\mu
\pi\right]+\ldots\ ,\cr
A_\mu =&\half i \left(\xi\partial^\mu\xi^\dagger -
\xi^\dagger\partial^\mu\xi\right) = {1\over f} \partial_\mu \pi + \ldots \ ,\cr
}}
that transform under $\sulr$ as $V_\mu\rightarrow U V_\mu U^\dagger +
\partial_\mu U U^\dagger$, and $A_\mu\rightarrow U A_\mu U^\dagger$. The
leading order baryon Lagrangian is
\eqn\blag{
\CL = \Tr \bar B \,i\,\Dsl\, B - m_B \Tr \bar B B - D \Tr \bar B
\gamma^\mu\gamma_5
\{A^\mu,B\} -  F \Tr \bar B \gamma^\mu\gamma_5 [A^\mu,B] \ ,
}
where ${\rm D}=\partial+V$ is a covariant derivative, $m_B$ is the $SU(3)$
invariant average mass of the octet baryons, and $D$ and $F$ are the usual
pion-nucleon axial coupling constants. This Lagrangian contains a factor of
$m_B$, which is dimensionful. The existence of $m_B$ destroys all the
dimensional analysis arguments used in the purely Goldstone boson sector. In
particular, one can have loop corrections of order $m_B/\lchi$ (which is of
order one), so that radiative corrections are no longer small, and the entire
calculational scheme breaks down. There must exist an effective theory for the
interaction of baryons with low momentum pions, independent of the value of the
baryon mass $m_B$. This is true because baryon number is a conserved quantity,
and so the overall mass of the baryon can be factored out of the problem. What
is relevant is not the energy of the baryon, but whether it is off-shell by an
amount small compared to $\lchi$\ref\jm{E. Jenkins and A. Manohar,
\pl{255}{1991}{558}}. Thus there must exist an effective field theory
describing the interactions of nearly on-shell baryons with low momentum pions.
This idea generalizes to other matter fields which carry a quantum number that
is conserved by the strong interactions, \eg\ it applies to baryons and mesons
containing a heavy quark. It cannot, however, be applied to $\rho$ mesons,
because a $\rho$ meson can decay into two pions with $q^2$ of order $m_\rho^2$,
so that an effective low energy description is not valid.

The theory of nearly on-shell baryons interacting with pions can be formulated
in terms of velocity dependent baryon fields,${}^{\jm}$ similar to the velocity
dependent quark fields introduced in the heavy quark effective
theory\ref\georgi{H.~Georgi, \pl{240}{1990}{447}}. The velocity dependent
baryon field $B_v$ is related to the baryon field $B$ by
\eqn\bvfield{
B_v(x) = e^{i m_B v \cdot x} \ {1+\slash v\over 2}B(x)\ .
}
In the rest frame $v=(1,0,0,0)$, the field $B_v$ has only the particle
components, and has the rest energy $m_B$ factored out. The chiral Lagrangian
eq.~\blag\ written in terms of the velocity dependent baryon fields
is${}^{\jm}$
\eqn\bvlag{
\CL_1 = \Tr \bar B_v i (v\cdot {\rm D}) B_v  - D \Tr \bar B_v
\gamma^\mu\gamma_5
\{A^\mu,B_v\} -  F \Tr \bar B_v \gamma^\mu\gamma_5 [A^\mu,B_v].
}
Note that the baryon mass term $m_B\bar B B$ is no longer present in the
Lagrangian. Thus the dimensional analysis arguments in the meson sector apply,
and one has a well defined theory with which to compute loop corrections. The
velocity dependent formalism is an expansion about the $m_B=\infty$ limit. One
can include $1/m_B$ corrections by including higher dimension operators into
the effective theory in a systematic way.
The baryon-pion interactions start at order $p^1$, and the baryon propagator is
$i/k\cdot v$. This changes the power counting rules from eq.~\power\ to
\eqn\bpower{
n -1 =\sum_{k\ge 2} m_k \left(k - 2\right) + \sum_{r\ge 1} n_r\left(r-1\right)
+ 2 L \ ,
}
for a graph containing one baryon line, $L$ loops,  $m_k$ insertions of the
meson interactions from $\CL_k$ and $n_r$ insertions of the baryon interactions
from $\CL_r$. The baryon chiral Lagrangian eq.~\bvlag\ has been used to compute
the leading non-analytic corrections to the baryon axial currents\ref\axial{E.
Jenkins and A. Manohar, \pl{259}{1991}{353}}, masses\ref\masses{E. Jenkins,
\np{368}{1992}{190}}, non-leptonic decays\ref\nonlep{E. Jenkins,
\np{375}{1992}{561}}, magnetic moments\ref\moment{E.~Jenkins, M.~Luke,
A.~Manohar, and M.~Savage, \pl{302}{1993}{482}}, weak radiative
decays\ref\wrhd{E.~Jenkins, M.~Luke, A.~Manohar, and M.~Savage, CERN preprint
CERN-TH-6690}, the sigma term\ref\sigmaterm{E. Jenkins and A. Manohar,
\pl{281}{1992}{336}},  electromagnetic polarizibilities\ref\empol{M.N. Butler
and M. Savage, \pl{294}{1992}{369}} and electromagnetic hyperon
decays\ref\emhd{M.N. Butler, M.J. Savage, and R.P. Springer,
\pl{304}{1993}{353}, UCSD-PTH-92-37}.
There has also been other extensive work on baryon chiral
perturbation theory.${}^{\gl}$\ref\swein{S. Weinberg, \pl{251}{1990}{288}\semi
S. Weinberg, \np{363}{1991}{3}\semi S. Weinberg,
\pl{295}{1992}{114}}\ref\bern{V. Bernard, N. Kaiser, and U. Meissner,
BUTP-93-09\semi V. Bernard, N. Kaiser, and U. Meissner, BUTP-93-05\semi V.
Bernard, N. Kaiser, and U. Meissner, \prl{69}{1992}{1877}\semi For a review,
see U. Meissner, BUTP-93-01}

One can also apply the velocity dependent formalism to mesons containing a
heavy quark, such as the $B$ and $B^*$ mesons by introducing velocity dependent
fields $B_v$ and $B_v^{*\mu}$. It is convenient to combine the two fields into
a single field
\eqn\hfield{
H_v = {1+\slash v\over 2} \left[ {\hbox{/\kern-.7000em $B^*_v$}}
 - B_v
\gamma_5\right]\ ,
}
which transforms under the heavy quark symmetry and chiral symmetry as
\eqn\htrans{
H \rightarrow S H U^\dagger \ .
}
The field $H_v$ can then be used to write down an effective Lagrangian for the
interactions of $B$ mesons with pions which also respects heavy quark
symmetries\ref\wyd{M.B. Wise,
\physrev{D45}{1992}{2188}\semi G. Burdman and J. Donoghue,
\pl{280}{1992}{287}\semi T.M. Yan, \etal, \physrev{D46}{1992}{1148}}. The
leading terms in the effective theory are
\eqn\hlag{
\CL_1= - i  \Tr H (v\cdot {\rm D}) \bar H
  - g \Tr H A_\mu \gamma^\mu \gamma_5 \bar H\ .
}
The chiral Lagrangian for baryons containing a heavy quark has been used to
compute corrections to $f_D$ and $B-\bar B$ mixing\ref\grinstein{B.~Grinstein,
et al., \np{380}{1992}{369}}, the Isgur-Wise function\ref\js{E.~Jenkins and
M.~Savage, \pl{281}{1992}{361}}, radiative decays\ref\raddec{J. Amundson, et
al., \pl{296}{1992}{415}\semi P. Cho and H. Georgi, \pl{296}{1992}{408}\semi
H.-Y. Cheng et al., \physrev{D47}{1993}{1030}}, masses\ref\bdmass{J.L. Goity,
\physrev{D46}{1992}{3929}\semi E. Sather and L. Randall, MIT-CTP-2166\semi E.
Jenkins, CERN-TH-6765}, and non-leptonic decays\ref\bdnon{H.-Y. Cheng et al.,
\physrev{D46}{1992}{5060}}. Some of these topics are covered by the other
speakers at this workshop.

\footatend\vfill\supereject\immediate\closeout\rfile\writestoppt
\baselineskip=0.5truecm\centerline{REFERENCES}\bigskip{\frenchspacing%
\parindent=20pt\escapechar=` \input refs.tmp\vfill\eject}\nonfrenchspacing
\bye